\documentclass[aps,pra,twocolumn,superscriptaddress,showpacs,10pt]{revtex4-1}

\usepackage{graphicx}
\usepackage{epstopdf}
\usepackage{amsmath}
\usepackage{verbatim} 
\usepackage{color}   
\usepackage{subfigure}
\usepackage{amssymb}
\usepackage{amsmath}
\usepackage{mathrsfs}
\usepackage{amsfonts}
\usepackage{braket}
\usepackage{appendix}
\DeclareMathAlphabet{\mathpzc}{OT1}{pzc}{m}{it}

\newcommand{\bob}[1]{
_\mathrm{#1}
}

\newcommand{\colvec}[3]{
    \renewcommand{\arraystretch}{1}
    \left(
    \!
    \begin{array}{c}
    #1 \\ #2 \\ #3
    \end{array}
    \!
    \right)
}

\begin{document}

\title{Dynamics of a single trapped ion immersed in a buffer gas}

\author{Bastian H{\"o}ltkemeier}
\author{Pascal Weckesser} \thanks{Currently at Physikalisches Institut, Albert-Ludwigs-Universit{\"a}t Freiburg}
\author{Henry L\'{o}pez-Carrera} 
\affiliation{Physikalisches Institut, Ruprecht-Karls-Universit{\"a}t Heidelberg, INF 226, 69120 Heidelberg, Germany}
\author{Matthias Weidem{\"u}ller}  \thanks{weidemueller@uni-heidelberg.de}
\affiliation{Physikalisches Institut, Ruprecht-Karls-Universit{\"a}t Heidelberg, INF 226, 69120 Heidelberg, Germany}
\affiliation{Hefei National Laboratory for Physical Sciences at the Microscale and Department of Modern Physics, and CAS Center for Excellence and Synergetic Innovation Center in Quantum Information and Quantum Physics, University of Science and Technology of China, Hefei, Anhui 230026, China.}

\date{\today}

\begin{abstract}

We provide a comprehensive theoretical framework for describing the dynamics of a single trapped ion interacting with a neutral buffer gas, thus extending our previous studies on buffer-gas cooling of ions beyond the critical mass ratio [B. H\"oltkemeier \emph{et al.}, Phys. Rev. Lett. \textbf{116}, 233003 (2016)]. By transforming the collisional processes into a frame, where the ion's micromotion is assigned to the buffer gas atoms, our model allows one to investigate the influence of non-homogeneous buffer gas configurations as well as higher multipole orders of the radio-frequency trap in great detail. Depending on the neutral-to-ion mass ratio, three regimes of sympathetic cooling are identified which are characterized by the form of the ion's energy distribution in equilibrium. We provide analytic expressions and numerical simulations of the ion's energy distribution, spatial profile and cooling rates for these different regimes. Based on these findings, a method for actively decreasing the ion's energy by reducing the spatial expansion of the buffer gas arises (Forced Sympathetic Cooling).

\end{abstract}


\maketitle

\section{Introduction}
\label{intro}

The motion of an ion inside a radio frequency (rf) trap is characterized by the interplay between a fast oscillation driven by the rf field (micromotion) and a much slower bound motion in the confining ponderomotive potential (macromotion) \cite{Dehmelt1967, Prestage1991, Ghosh1996}.
If the ion is exposed to collisions with a neutral buffer gas, the dynamics becomes more complex.
Elastic collisions influence the permanent exchange of energy between micromotion and macromotion resulting in an efficient energy transfer from the micromotion to the macromotion. This may lead to an increase of the ion's energy, known as collisional heating, even if the buffer gas temperature is much lower than the ion's mean energy \cite{Cetina2012}. Similar processes are known as Intrabeam Scattering in storage rings of charged particles \cite{Piwinsky1987}.
With the recent advances in simultaneously trapping ions and ultracold atomic gases \cite{hart2014, Willitsch2014}, the possibilities of sympathetic cooling have been investigated in great detail \cite{Asvany2009, Devoe2009, Zipkes2011, Cetina2012, Chen2014}.
It was shown that for atom-to-ion mass ratios $\xi$ close to unity, the ion's energy \cite{Asvany2009, Zipkes2011, Chen2014} as well as the spatial distributions \cite{Devoe2009} exhibit a power law behavior for a buffer gas well described by a Boltzmann distribution.
For larger mass ratios, these power law tail in the energy distribution lead to a diverging mean ion energy which finally results in the ion's loss from the trap, even if the buffer gas is at zero temperature.
This phenomenon was first described by Major and Dehmelt \cite{Dehmelt1968} who derived a critical mass ratio of $\xi_{\mathrm{crit}} = 1$ marking the transition to the unstable regime. Some of the recent studies have determined the critical mass ratio numerically \cite{Devoe2009, Zipkes2011} as well as analytically \cite{Chen2014} finding values slightly larger than Major and Dehmelt's original prediction.

In order to extend the stable regime to larger mass ratios, two approaches can be pursued.
Firstly, spatially confined buffer gases such as atoms stored in optical traps \cite{Hudson2009, Grier2009, Zipkes2010a, Ravi2012, Schmid2012, Rellergert2013, Harter2013, Willitsch2014, Dutta2015}, restrict collisions to the trap center where the micromotion is smallest, thus reducing the collision induced heating \cite{Gerlich1992, Zipkes2011, Ravi2012}.
Secondly, rf traps with higher multipole orders feature a reduced micromotion in the center of the trap thus also reducing the effective energy increase through collisions \cite{Gerlich1992, Wester2009}.
Despite a growing number of experiments using these approaches, all previous theoretical studies are limited to homogeneous buffer gas distributions \cite{Dehmelt1968, Asvany2009, Devoe2009, Zipkes2011, Cetina2012, Chen2014}. In addition, most of them are only valid for quadrupole (Paul) rf traps \cite{Dehmelt1968, Devoe2009, Zipkes2011, Cetina2012, Chen2014}.

In a recent Letter \cite{prl2016} we extended the theoretical description of sympathetic cooling in rf traps to include localized buffer gas configurations and higher multipole orders of the rf-trap.  The ion's motion is described in the adiabatic approximation which, in contrast to the solution of the Mathieu equations allows to treat higher multipole orders. Elastic collisions with a neutral buffer gas are included by a frame transformation into the rest-frame, where the micromotion is assigned to the neutral buffer gas instead of the ion. In this frame, the buffer gas atoms have an effective spatially-dependent velocity. This transformation provides an intuitive understanding of collisional heating and explains the origin of a new stable regime at large mass ratios. Using numerical simulations, we find three distinct dynamical regimes, characterized by distinct analytical expressions for the ion's equilibrium energy distribution. These results not only comprise earlier studies on collisional cooling of ions but also predict a novel regime of stable cooling of ions beyond the critical mass ratio. As an additional outcome of these investigations, one can actively tune the ions temperature by controlling the buffer gas' extension and/or the rf-trapping fields (forced sympathetic cooling). In this paper we present a detailed description of our model and the numerical algorithms used to calculate the ion's final energy distributions.
We discuss the ion's equilibrium state in the different regimes, including the ion's spatial distributions and average cooling rates for different mass ratios and trap orders.

\section{Elastic buffer-gas collisions inside a radio-frequency ion trap}
\begin{figure}[t]
\includegraphics[width=\columnwidth, trim= 0.3cm 0.2cm 0.3cm 0.1cm]{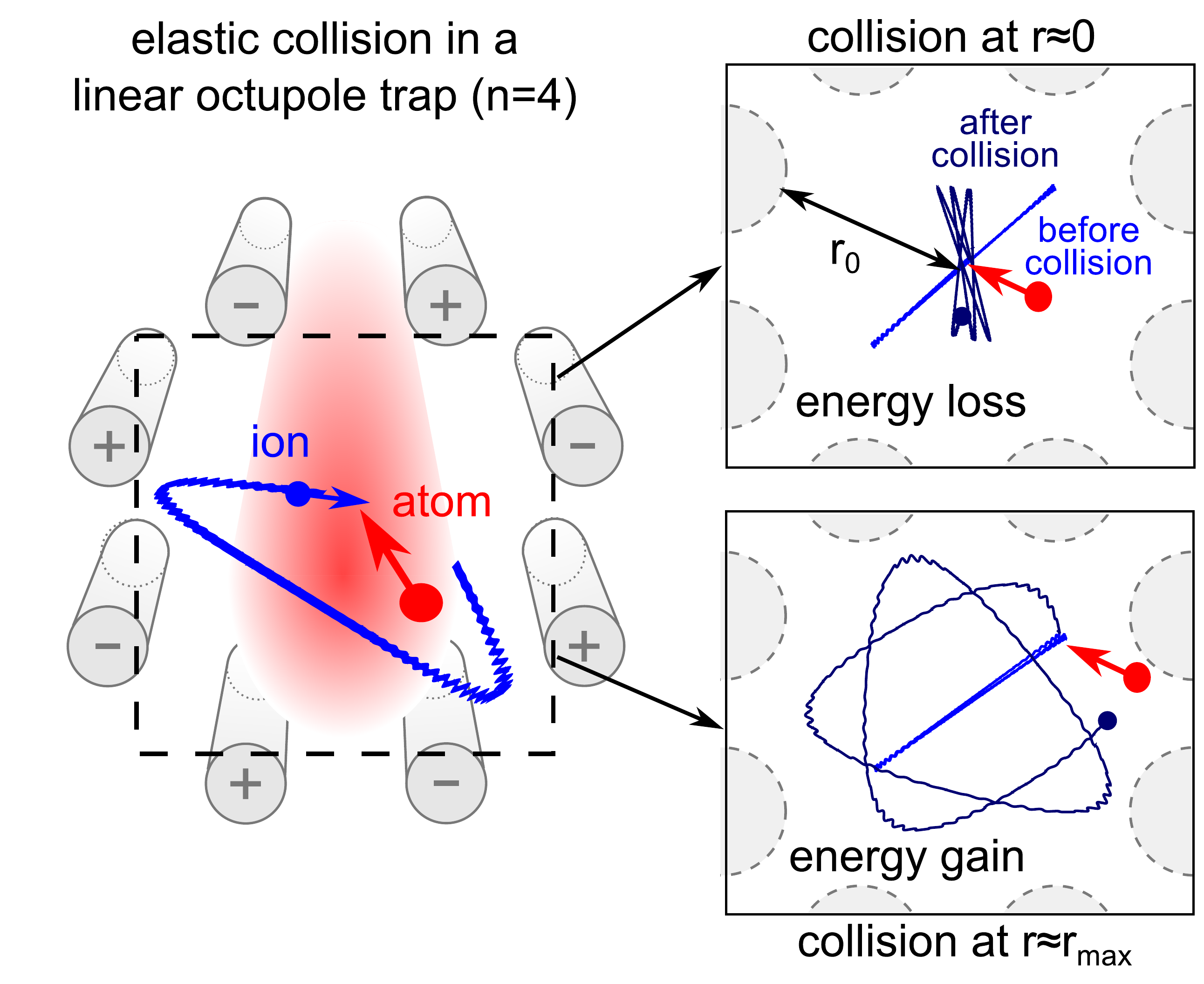}
\caption{
Schematic of a linear rf-trap with $n=4$. The atoms are indicated in red, the ion is shown in blue. 
The eight rf-electrodes are indicated by the gray cylinders.
Such kind of rods are commonly used to replace the ideal hyperbolic electrodes \cite{Gerlich1992}.
The two insets on the right illustrate two elastic collisions, one close to the trap center and the other close to the ions turning point $r_\mathrm{turn}$.
The trap radius $r_0$ is indicated by the arrow.
}
\label{fig:NewFigure}
\end{figure}

\subsection{Ion motion in a radio-frequency trap}
Multipole radio-frequency (rf) traps provide an effective way to spatially confine charged particles.
An ideal linear multipole trap of order $n$ creates a two dimensional electric field with only the $n$'th term of the multipole expansion being non-zero \cite{Gerlich1992}.
This can be achieved by applying a voltage of ${\pm U_0}$ to $2n$ hyperbolically shaped electrodes (see Fig. \ref{fig:NewFigure}).
Choosing a voltage of ${U_0 = U \cos (\omega t)}$ and the distance of two adjacent electrodes to be $2 r_0$ results in a force of
\begin{equation}
\vec{F}(r, \phi,t) = \frac{Q_\mathrm{i} U n}{r_0^{n}} r^{n-1} \, \cos(\omega t) \, \colvec{-\cos{}[(n-1)\phi]}{\ \ \ \sin{} [(n-1)\phi]}{0}
\label{trapDGL}
\end{equation}
acting on an ion placed inside the trap, with $Q\bob{i}$ and $m \bob{i}$ being the ion's charge and mass and ${r=\sqrt{x^2+y^2}}$ and ${\phi=\arctan{}\!(y/x)}$ being cylindercal coordinates with the $z$-axis being parallel to the electrodes.
The unit vector of the multipole field at the end of Eq.~\ref{trapDGL} will be denoted by $\vec{\mathrm{e}}_\mathrm{mp}$ in the following.
For the special case $n=2$ (Paul trap) these equations decouple and are equivalent to the Mathieu differential equations which can be solved analytically \cite{Mclachlan1951, Meixner2013}.
The solutions of the Mathieu equations are characterized by the AC-stability parameter ${q=(2 n Q\bob{i} U )/(m\bob{i} r_0^2 \omega^2)}$.

For all higher multipole orders ($n>2$) the equations of motion are coupled and Eq.~\eqref{trapDGL} is a nonlinear second order differential equation which cannot be solved analytically.
In this case an approximate solution can be found by separating the ion's motion $\vec{R}(t)$ into two distinct time scales \cite{Gerlich1992, Wester2009}.
The fast time scale is set by the ion's oscillation in the rf field $\vec{R}_\mathrm{rf}(t)$ (micromotion) and the slow time scale is given by the drift motion around the trap center $\vec{R}_\mathrm{d}(t)$ (macromotion).
Assuming that at all times $\vec{R}_\mathrm{rf}(t) \ll \vec{R}_\mathrm{d}(t)$ and that the characteristic time scale of the macromotion is much longer than an rf period, Eq.~\eqref{trapDGL} can be expanded in a Taylor series and the ion's equations of motion on both time scales can be solved separately.
The validity of this so called adiabatic approximation is expressed by the multipole stability parameter \cite{Teloy1974, Gerlich1992}
\begin{equation}
\eta= q (n\! -\! 1) \left( \frac{r}{r_0} \right)^{n-2} \; ,
\label{eq:stabParameter}
\end{equation}
with $q$ being the AC-stability parameter of the Matthieu equations, introduced earlier.

For $\eta \ll 1$ all assumptions of the adiabatic approximation are met and the velocity of the micromotion at any position inside the trap is given by
\begin{equation}
\vec{v}_\mathrm{rf}(r,\phi,t) =  \frac{\omega q r_0}{2} \left( \frac{r}{r_0} \right)^{n-1} \,  \sin(\omega t) \, \, \vec{\mathrm{e}}_\mathrm{mp} \; .
\label{eq:Rrf}
\end{equation}
The macromotion can be expressed as a motion in the effective potential
\begin{equation}
V_\mathrm{eff} (r)=\frac{q^2 \omega^2}{16} \left( \frac{r}{r_0} \right)^{2n-2} \; .
\label{eq:effectivePot}
\end{equation}
which is equivalent to the kinetic energy of the micromotion, averaged over one rf oscillation.

\newcommand{\mimoD}[0]{\vec{v}_\mathrm{rf} \! \! \! ' \ }
\newcommand{\mamoD}[0]{\vec{v}_\mathrm{d \, } \! \! \! ' \ }
\newcommand{\atomD}[0]{\vec{u}_\mathrm{a} \! \! \! ' \ }
\newcommand{\ionD}[0]{\vec{v}_\mathrm{i \, } \! \! ' \, }

\subsection{Buffer gas collisions in the presence of micromotion}

\begin{figure*}[bht]
\includegraphics[width=\textwidth, trim= 0.3cm 0.2cm 0.3cm 0.1cm]{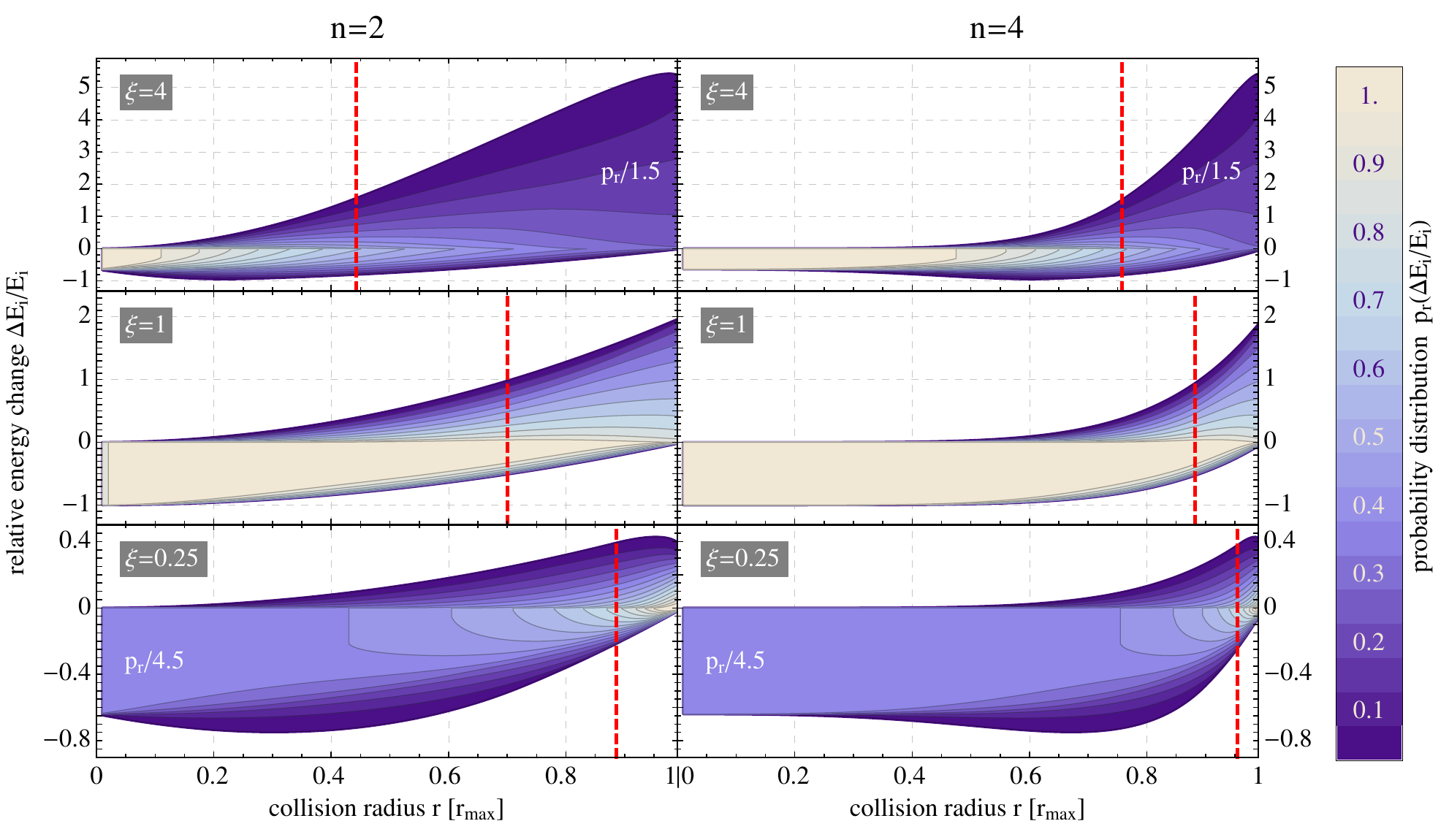}
\caption{
Probability distributions of a relative energy change $\Delta E_i/E_i$ in an elastic ion-neutral collision for an ion at energy $E_i$ as a function of the normalized collision radius $r/r_\mathrm{max}$. The buffer gas is assumed to be at rest (temperature $T_a=0$). Left graphs show the distribution for a Paul trap (pole order $n=2$), right graps for an octupole trap ($n=4$). The lower, middle and upper graphs correspond to mass ratios of $\xi$ = 0.25, 1 and 4, respectively. The color code provides the normalized probability function $p_r(\Delta E_i /E_i)$ of the energy change at a given radius r. Indicated by the red dashed line is the critical cooling radius (Eq.~\eqref{eq:R_cool}) given by the condition that average energy transfer $\left< \Delta E_i / E_i \right>$ becomes zero.
}
\label{fig0}
\end{figure*}

We will show that, based on the adiabatic approximation, one can find analytic expressions for the change of the ion's energy due to an elastic collision with a neutral buffer gas atom. The ion's total kinetic energy oscillates with the rf field and is therefore not a good measure for the state of the system. A more suitable definition is obtained by replacing the kinetic energy of the micromotion by its average value, namely the effective potential. This results in a constant value of the ion's energy $E_i$ as given by
\begin{equation}
E_i = \frac{1}{2} m_i v_d^2 + V_\mathrm{eff}(r) \, ,
\label{eq:Er}
\end{equation}
with $v_d$ being the velocity of the macromotion.

Consequently, the change of the ion's energy due to an elastic collision (see Fig.~\ref{fig:NewFigure}) is equivalent to the change of the kinetic energy of the macromotion, i.e. $\Delta E_i = \frac{1}{2} m_i (\mamoD ^2 -\vec{v}_d \, \! ^2)$.
The velocity of the macromotion after the collision $\mamoD$ is calculated by a transformation into the center-of-mass (COM) frame which is defined by the COM velocity $\vec{V}_\mathrm{COM}= \frac{m_{\mathrm{i}}\vec{v}_{\mathrm{i}}+m_{\mathrm{a}}\vec{u}_{\mathrm{a}}}{m_{\mathrm{i}}+m_{\mathrm{a}}}$, with $\vec{v}_{\mathrm{i}} = \vec{v}_{\mathrm{d}} + \vec{v}_{\mathrm{rf}}$ being the ion's total velocity and $\vec{u}_{\mathrm{a}}$ being the buffer gas atom's velocity.
The final ion velocity reads 

\begin{equation}
\ionD =\vec{V}_\mathrm{COM}+\frac{m_{\mathrm{a}}}{m_{\mathrm{i}}+m_{\mathrm{a}}} \mathcal{R}(\theta_c,\phi_c) (\vec{v}_{\mathrm{i}}-\vec{u}_{\mathrm{a}})  \; ,
\label{eq:finalVelo1}
\end{equation}

with $\mathcal{R}(\theta_c,\phi_c)$ being a rotation matrix with the polar and azimuthal scattering angles $\theta_\mathrm{c}$ and $\phi_\mathrm{c}$.

From this expression, the final velocity of the macromotion is obtained by decomposing the ion's total velocity before and after the collision into its two components, namely ${\vec{v}_\mathrm{i} = \vec{v}_\mathrm{d} + \vec{v}_\mathrm{rf}}$ and ${\ionD = \mamoD + \mimoD}$.
Assuming that the duration of the collision is short compared to the period of the micromotion, the micromotion velocity, which is determined by the ion's position in the rf field, remains unchanged through the collision ($\vec{v}_\mathrm{rf} = \mimoD$).
Eq.~\eqref{eq:finalVelo1} then yields

\begin{equation}
\mamoD =  \frac{m_{\mathrm{i}}\vec{v}_{\mathrm{d}}+m_{\mathrm{a}}\vec{u}_\mathrm{eff}}{m_{\mathrm{i}}+m_{\mathrm{a}}} + \frac{m_{\mathrm{a}}}{m_{\mathrm{i}}+m_{\mathrm{a}}}\mathcal{R}(\theta_c,\phi_c) (\vec{v}_{\mathrm{d}}-\vec{u}_\mathrm{eff}) \; ,
\label{eq:finalVelo2}
\end{equation}
with $\vec{u}_\mathrm{eff} = \vec{u}_\mathrm{a}-\vec{v}_{\mathrm{rf}}$ being the effective atom velocity.

The effective atom velocity is chosen in a way that Eq.~\eqref{eq:finalVelo1} and \eqref{eq:finalVelo2} become formally equivalent with $\vec{v}_\mathrm{i} \simeq \vec{v}_\mathrm{d}$ and $\vec{u}_\mathrm{a} \simeq \vec{u}_\mathrm{eff}$.
Consequently, elastic collisions in an rf-trap can be described by an ion moving in the effective potential $V_\mathrm{eff}$, colliding with buffer gas atoms exhibiting the spatial and time dependent velocity $\vec{u}_\mathrm{eff}$.
In this picture, the average effective energy of the buffer gas atoms is given by

\begin{equation}
\left< E_a \right> = \frac{1}{2}m_a \left( \left< \vec{u}_a^2 \right> + 2 \left< \vec{u}_a \vec{v}_\mathrm{rf} \right> + \left< \vec{v}_{\mathrm{rf}}^2 \right> \right) \, .
\label{eq:effectiveEnergy}
\end{equation}

The first term on the right side is the thermal energy of the buffer gas, the second term vanishes, as the thermal velocity of the atoms is not correlated with the effective micromotion, and the last term is proportional to the ponderomotive potential.
For a thermal buffer gas of temperature $T_a$ this results in an average energy of

\begin{equation}
\left< E_{\mathrm{a}} \right> = \xi V_\mathrm{eff}(r) + \frac{3}{2} k_\mathrm{B} T_\mathrm{a} \, ,
\label{eq:effectiveEnergy}
\end{equation}

with $k_B$ being the Boltzmann constant.
The buffer gas thus exhibits an effective energy distribution given by a Boltzmann distribution at the center of the trap, where micromotion can be neglected, with an increasing contribution of the micromotion for increasing radii.

By comparing the radial dependence of the atoms effective energy to the ion energy $E_i$, as given by $\frac{1}{2}m_i v_d^2(r)$ (see Eq.~\eqref{eq:Er}), two distinct regions, separated by the critical cooling radius $r_\mathrm{c}$, can be identified: For $r<r_c$ the ion's kinetic energy exceeds the average energy of the atoms resulting in a net energy transfer to the atoms, whereas for $r>r_\mathrm{c}$ the ion's energy is generally increased through a collision with the buffer gas atom.
The radius $r_\mathrm{c}$ is obtained by solving the equation $\left< E_{\mathrm{a}} (r_c) \right> = E_i - V_\mathrm{eff}(r_c)$ which leads to a critical cooling radius of

\begin{equation}
r_{\mathrm{c}} =r_{\mathrm{max}} \left( \frac{1- \frac{3}{2} k_\mathrm{B} T_\mathrm{a}/E_\mathrm{i}}{1+\xi} \right)^{\frac{1}{2n-2}} \ ,
\label{eq:R_cool}
\end{equation}

with $r_\mathrm{max}$ being the ion's maximum turning point in the ponderomotive potential, as defined by the condition $V_{\mathrm{eff}}(r_\mathrm{max})=E_\mathrm{i}$.
For $E_\mathrm{i}< k_\mathrm{B} T_\mathrm{a}$, the net energy transfer is always positive and, thus, the radius $r_\mathrm{c}$ no longer defined.
In case the ion's energy is large compared to $\frac{3}{2} k_B T_a$, the cooling volume is always a fixed fraction of the total volume probed by the ion.

The probability of a relative energy change $\Delta E_i/E_i$ to occur at different collision radii is shown in Fig.~\ref{fig0} for the case ($E_i \gg \frac{3}{2} k_B T_a$). The normalized distribution function $p_r(\Delta E_i/E_i)$ gives the probability for a collision at radius $r$ to result in a relative energy change of $\Delta E_i/E_i$. The derivation of the the analytical expressions for these probability distribution will be published elsewhere. In combination with the probability distribution $p_i(r)$ to find the ion at radius $r$, this yields the total probability distribution
\begin{equation}
P(\Delta E_i /E_i) = \int p_r(\Delta E_i /E_i) p_i(r) \, \mathrm{d}r
\end{equation}
for an energy change of $\Delta E_i /E_i$ to occur. Expressions for radial probability $p_i(r)$ will be derived in the next section.

Fig.~\ref{fig0} illustrates that collisions close to the trap center always reduce the ion's energy whereas collisions close to $r_\mathrm{max}$ increase the energy, with the two regimes being separated by the critical cooling radius $r_c$ indicated by the red dashed line defined by $\langle \Delta E_i \rangle = 0$. As expected, the overall energy change also shows a pronounced dependence on the multipole order $n$ and the mass ratio $\xi$.
On the one hand, with growing multipole order the volume of efficient cooling at small radii is extended as the radial dependence of $p_r(\Delta E_i/E_i)$ scales as $(r/r_\mathrm{max})^{n-1}$.
On the other hand, the micromotion induced heating at large radii is amplified with larger atom-to-ion mass ratio, as the velocity of the micromotion $v_\mathrm{rf} \propto 1/m_i$, now being assigned to the atom of mass $m_a$, leads to the prefactor $\xi$ in the effective buffer gas energy (see Eq.~\eqref{eq:effectiveEnergy}).
Hence, at large mass ratios and small multipole orders, the ion essentially gains energy through collisions, whereas at low mass ratios and large multipole orders the energy is primarily reduced.

These consideration offer an intuitive picture for the occurence of a critical mass ratio $\xi_\mathrm{crit}$, beyond which the ion effectively gains energy through buffer-gas collisions, eventually resulting the ion's loss from the trap. The critical mass ratio can be estimated by averaging the energy change ${\Delta E_i = \frac{1}{2} m_i (\mamoD ^2 -\vec{v}_\mathrm{d} \, \! ^2)}$ over one rf-period which results in
\begin{equation}
\left< \Delta E_{\mathrm{i}} \right> = \frac{m_i m_a (1-\cos{\theta_c})}{(m_i + m_a)^2} \left[ m_a \left< v_\mathrm{rf}^2 \right> - m_i \left< v_\mathrm{d}^2 \right> \right] .
\label{eq:delEav}
\end{equation}
The first term on the right side is proportional to the average effective potential $\left< V_\mathrm{eff} \right>$, the second term to the average kinetic energy of the macromotion $\left< E_d \right>$.
For the critical mass ratio, the average energy transfer has to be zero, which leads to the condition ${\left< E_d \right> = \xi_\mathrm{crit} \left< V_\mathrm{eff} \right>}$.
Assuming that the ion motion in the trap's effective potential obeys the virial theorem, the ratio between average potential and kinetic energy is given by ${\langle V_{\mathrm{eff}} \rangle / \langle E_\mathrm{d} \rangle = 2/(3n-3)}$ which results in a critical mass ratio of
\begin{equation}
\xi_\mathrm{crit} = \frac{3}{2} (n-1) \, .
\label{eq:xiCrit}
\end{equation}
The critical mass ratio grows linearly with the multipole order of the trap.
Consequently, for any mass ratio a sufficiently large multipole order can be found which leads to stable trapping conditions.
It should be noted, that Major and Dehmelt \cite{Dehmelt1968} originally estimated the critical mass ratio for a Paul trap to be $\xi_\mathrm{crit} = 1$.
This reduced critical mass ratio corresponds to a pure two dimensional system, where the ion has no velocity component in $z$-direction.
In the following we will numerically calculate the critical mass ratio as the point where the mean energy of the ion's energy distribution diverges \cite{Devoe2009, Zipkes2011, Chen2014}.
This alternative definition is commonly used in the literature and leads to the same linear multipole order dependence with a slightly varied pre-factor.

\section{Numerical simulations}

In order to numerically determine the ion's equilibrium energy distribution, the energy $E_\mathrm{i}$ has to be tracked over the course of many collisions. In previous investigations, this was done using the solutions of the Mathieu differential equations \cite{Devoe2009, Zipkes2011, Chen2014} or full trajectory calculations \cite{Asvany2009, Ravi2012, lopez2015}. The first approach has the disadvantage, that it is limited to Paul traps and homogeneous buffer gases. The second approach lifts these limitations as the full trajectory calculation can be applied to traps of any multipole order and evaluating the collision probability $P(t) \mathrm{d}t$ for every infinitesimal time step $\mathrm{d}t$ allows to simulate arbitrary buffer gas configurations.
However, full trajectory calculations result in very long computation times, as the time between consecutive collisions is usually much longer than the time scales set by micro- and macromotion.
In the following we develop a general method, which can be used for any multipole order and buffer gas configuration, without the necessity to calculate the ion's full trajectory.

\subsection{Numerical model}
\begin{figure}[t]
\includegraphics[width=\columnwidth, trim= 0.3cm 0.2cm 0.3cm 0.1cm]{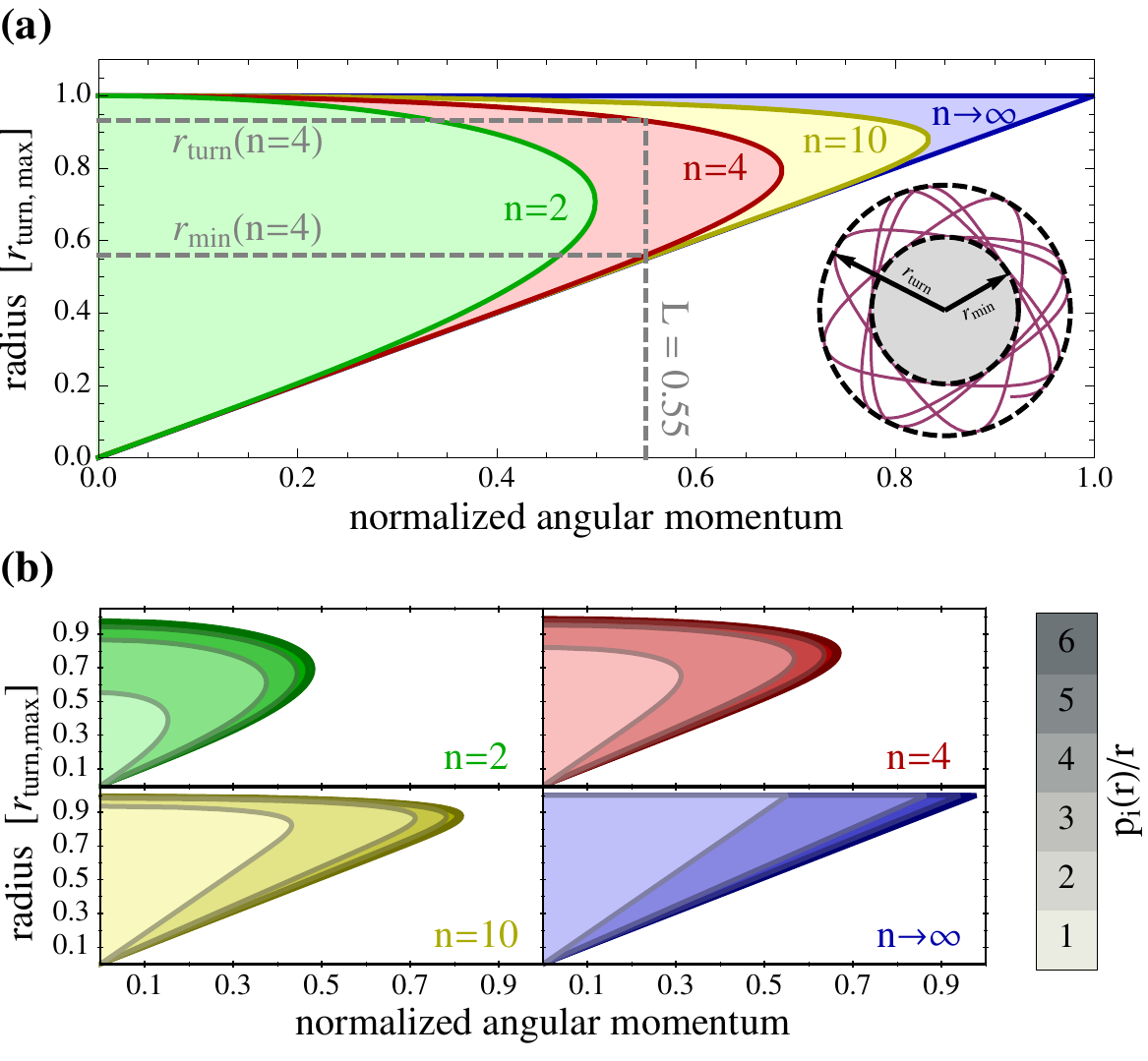}
\caption{
Ion's spatial probability distribution. The upper graph (a) shows the ion's minimum and maximum radius for different angular momenta and multipole orders. The angular momentum has been normalized using the maximum turning point $r_\mathrm{turn, max}$, given by the condition $V_\mathrm{eff}(r_\mathrm{turn, max})=E_\mathrm{r}$ resulting in ${L_\mathrm{norm}= \vert (\vec{r}/r_\mathrm{turn, max}) \times (\vec{v}_\mathrm{d,r}/v_\mathrm{d,r}(r=0)) \vert}$. The inset shows an exemplary trajectory (red) with $r_\mathrm{turn}$ and $r_\mathrm{min}$ being indicated by the dashed circles. The lower graph (b) shows the radial probability distribution as a function of the angular momentum. In analogy to a one-dimensional pendulum, the probability distribution diverges at the radial turning points as the radial velocity $\dot{r}$ becomes zero. 
}
\label{fig1}
\end{figure}
Our model is based on three main simplifying assumptions.
First, we use the adiabatic approximation to describe the motion of the ion.
This limits our model to small stability parameters $\eta$ (see Eq.~\eqref{eq:stabParameter}).
We find satisfactory agreement with the ion's exact trajectory for $\eta < 0.1$, which is also the regime most experiments are operated in. 
Second, for the energy regime discussed here, the classical Langevin model \cite{Langevin1905,Gioumousis1958} is used, which yields a velocity-independent scattering probability for ion-atom collisions.
This is valid, as long as a large number of partial waves contribute to the scattering process and quantum interferences of scattering amplitudes can be neglected.
Third, we assume that the time between consecutive collisions is long compared to the period of the macromotion, allowing one to apply a separation of time scales.
As long as the ion undergoes many oscillations in the effective potential (macromotion) between every two consecutive collisions, the ion's exact trajectory can be substituted by a radial density distribution $p_\mathrm{i}(r)$.
The density distribution is independent of the two other cylindrical coordinates $z$ and $\phi$, as we assume perfect axial symmetry.

Based on these three simplifications, the probability of an collision to occur in the interval $dr$ around a radial position $r$ is given by the relative overlap of the buffer gas and ion distributions
\begin{equation}
\bar{P}(r)\,dr = \rho_a(r) p_i(r) \,dr \, ,
\label{eq:delEav}
\end{equation}
with $\rho_a(r)$ being the normalized density distribution of the buffer gas.

The ion's spatial probability distribution $p_i(r)$ is characterized by the ion's total energy and angular momentum.
The total energy can be separated into two components, the energy $E_\mathrm{z} = m_i v_\mathrm{d,z}^2 /2$ in axial direction without confining potential and the radial energy $E_\mathrm{r} = m_\mathrm{i} v_\mathrm{d,r}^2 /2 + V_\mathrm{eff}(r)$ consisting of macromotion and ponderomotive potential, with $v_\mathrm{d,r}$ and $v_\mathrm{d,z}$ being the radial and axial components of the macromotion.
Together with the ion's angular momentum $L=m_i \vert \vec{r} \times \vec{v}_\mathrm{d,r} \vert$, these quantities are constants of the ion's motion and uniquely define $p_\mathrm{i}(r)$.
The probability distribution is proportional to the inverse of the ion's radial velocity $\dot{r}$, which can be expressed as
\begin{equation}
p_i(r) \propto r \left(\frac{2}{m_i} (E_\mathrm{r}-V_\mathrm{eff}(r)) - \frac{L^2}{r^2} \right)^{-1/2} \, .
\label{eq:piofr}
\end{equation}
Fig.~\ref{fig1} shows the resulting radial probability distribution $p_i(r)/r$ for different multipole orders.

In the simulation, the ion's radial and axial energy as well as it's angular momentum are computed and stored after every collision.
As initial conditions we use $E_\mathrm{r}=k_\mathrm{B} T_\mathrm{a}$, $E_\mathrm{z}=k_\mathrm{B} T_\mathrm{a} /2$ and $L=0$. As the simulations comprise a large number of collisions (typically one million), the final distributions become independent of the choice of the initial conditions.
For every collision we then pick the following set of parameters:
\begin{itemize}
\item \textbf{Rf-phase} - the phase of the rf-field $\phi_\mathrm{rf}$ is randomly chosen between 0 and 2$\pi$.
\item \textbf{Atom velocity} - all three cartesian coordinates of the atom's velocity vector are chosen with a normal distribution with standard deviation $\sqrt{k_\mathrm{B} T_\mathrm{a}/m_\mathrm{a}}$.
\item \textbf{Collision radius} - the radial position of the collision $r_\mathrm{coll}$ is chosen according to $P(r)$ which is determined by $L$ and $E_r$ (see Eq.~\eqref{eq:piofr}). For large radial ion energies P(r) can have a long tail of near zero values, in which case we use an upper boundary of $r_\mathrm{coll} < 5 \sigma_\mathrm{a}$ with $\sigma_\mathrm{a}$ being the width of the buffer gas cloud.
\item \textbf{Incident angle} - the random incident angle $\theta_\mathrm{i}$ between the ion's macromotion and the atom's effective velocity determines the COM velocity. The angle is picked according to the Jakobian determinant $p(\theta_\mathrm{i})=\sin{\theta_\mathrm{i}}$.
\item \textbf{Scattering angles} - using the Langevin model, the scattering angle in the center-of-mass frame is distributed isotropically. This is achieved by randomly choosing an azimuthal angle $\phi_\mathrm{c}$ between zero and $2\pi$ and a polar angle $\theta_\mathrm{c}$ between zero and $\pi$ taking into account the Jakobian determinant $p(\theta_\mathrm{c})=\sin{\theta_\mathrm{c}}$.
\end{itemize}
Based on these parameters the ions micro- and macromotion velocities are calculated and using Eq.~\eqref{eq:finalVelo2}, the ion's macromotion after the collision is obtained.
The new macromotion velocity together with the collision radius defines the ion's radial and axial energy as well as it's angular momentum after the collision based on which the next collision parameters can be calculated.
The average time of free motion before the next collision occurs, is given by the inverse of the overlap of atom and ion distribution
\begin{equation}
\tau = \tau_\mathrm{coll}/\left( \int \rho_\mathrm{a}(r) p_i(r) \mathrm{d}r \right)\, .
\label{eq:tau}
\end{equation}
where $\tau_\mathrm{coll}$ is the collision time as derived from the Langevin cross section. After typically performing $10^6$ such collisions, the distributions have converged. All energy values are weighted with the corresponding  $\tau$ and binned, resulting in the ion's steady-state energy distribution.

 \section{Equilibrium regimes}
\begin{figure}[t]
\includegraphics[width=\columnwidth, trim= 0.3cm 0.2cm 0.3cm 0.1cm]{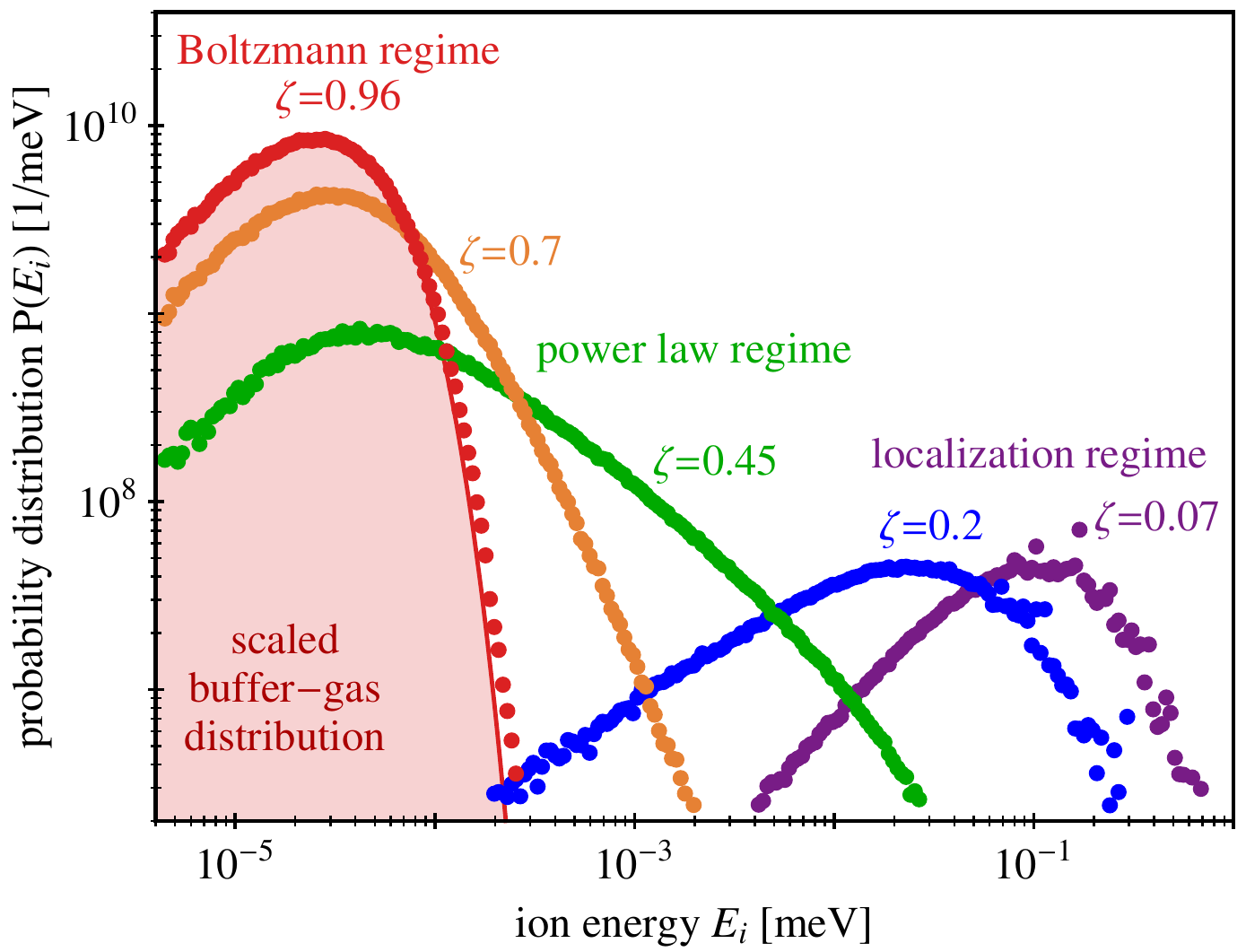}
\caption{
Normalized equilibrium energy distributions for different thermalicity parameters $\zeta$ in a Paul trap ($n=2$). The buffer gas cloud distribution is given by a Gaussian of size $\sigma_a=R_0/100$ and temperature of $T_a=200 \mu$K. Also shown is the energy distribution in the Boltzmann regime (red curve) and an energy distribution in the localization regime for $\xi=34$ (purple curve), according to Eq.~\eqref{eq:BoltzmannE} and \eqref{eq:LocalizationE}.
}
\label{fig2}
\end{figure}

The ion's final energy distribution depends on the one hand on the spatial distribution, temperature and mass of the buffer gas atoms and on the other hand on the trap parameters, namely the multipole order $n$ and the stability parameter $\eta$.
Fig.~\ref{fig2} shows some exemplary energy distributions $P(E_{\mathrm{i}})$ for different mass ratios in a Paul trap.

These energy distributions can be separated into different regimes which are characterized by the degree to which the effective buffer gas distribution in Eq.~\eqref{eq:PowerLawE} matches a thermal distribution ("thermalicity").
The atoms effective energy consists of two parts, the micromotion induced energy $\xi V_\mathrm{eff} (r)$ and the thermal energy of the buffer gas $\frac{3}{2}k_B T_a$.
So far we have only considered the radial dependence of the effective energy which resulted in the critical cooling radius (Eq.~\eqref{eq:R_cool}).
Additionally to this spatial dependence, the effective energy implicitly depends on the ion's energy $E_i$.
This dependence can expressed by replacing $V_\mathrm{eff}(r)$ by its average value given by the virial theorem, resulting in an effective energy of
\begin{equation}
\left< E_{\mathrm{a}} \right> = \frac{2 \xi}{3n-1} E_i + \frac{3}{2} k_\mathrm{B} T_\mathrm{a} \, .
\label{eq:effectiveEnergy2}
\end{equation}
A parameter $\zeta$ to determine the thermalicity is obtained by comparing the atoms' thermal energy $\frac{3}{2} k_\mathrm{B} T_\mathrm{a}$ to their total effective energy $\left< E_{\mathrm{a}} \right>$, assuming that the ion has an energy of $E_i = \frac{3}{2} k_\mathrm{B} T_\mathrm{a}$
\begin{equation}
\zeta = \frac{3n-1}{3n-1+2 \xi } \, ,
\label{eq:zeta}
\end{equation}
thus ranging from zero (distributions determined by the buffer gas' potential energy) to unity (distributions determined by the buffer gas' temperature). For $\zeta \approx 1$, the atom's effective energy distribution resembles a pure thermal distribution (Boltzmann regime).
With decreasing $\zeta$, the contribution of the micromotion induced energy becomes relevant and the ion's energy distribution exhibits growing power laws towards higher energies (power law regime). Finally for $\zeta$ approaching zero, stable energy distributions can only be realized using a localized buffer gas (localization regime).
These three regimes will be discussed in detail in the following.

\subsection{Boltzmann regime}

\begin{figure}[bt]
\includegraphics[width=\columnwidth, trim= 0.3cm 0.2cm 0.3cm 0.1cm]{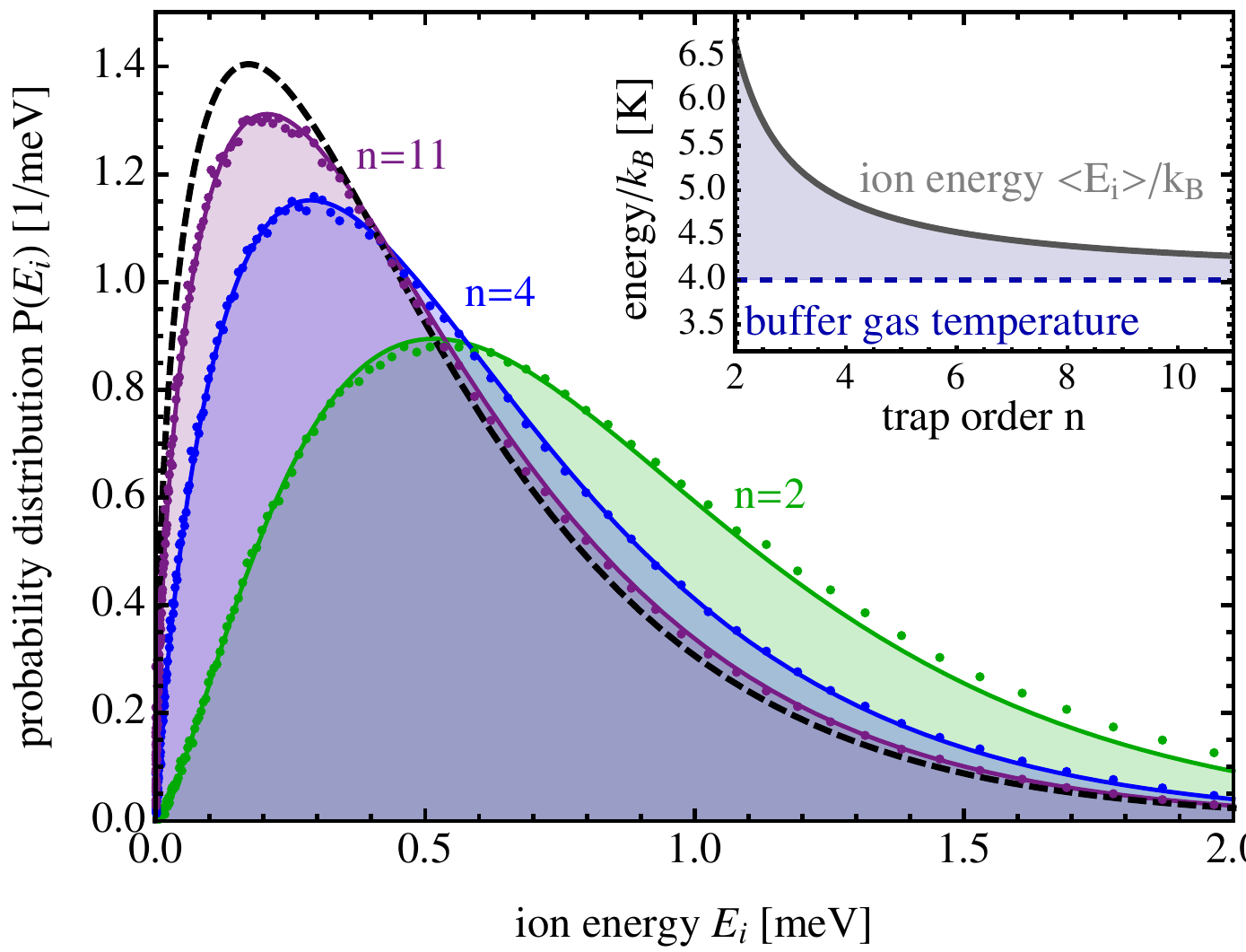}
\caption{
Ion's energy distribution in the Boltzmann regime.
The difference between the energy distribution of the ion and the one of the buffer gas (dashed line) is caused by the effective potential, confining the ion in the trap.
Depending on the multipole order, the influence of the effective potential differs.
The inset shows the ion's mean energy as a function of the trap order.
The mean energy is largest for an ion inside a Paul trap and converges towards the mean energy of the buffer gas for large multipole orders.
}
\label{fig3}
\end{figure}
For $\zeta \approx 1$ (which corresponds to $\xi \ll 1$) the ion thermalizes to the buffer gas temperature as the atoms' effective energy is dominated by the thermal energy (see Eq.~\eqref{eq:effectiveEnergy2}).

Fig.~\ref{fig3} shows the resulting energy distributions for three different multipole orders with ${\zeta = 0.95}$ as well as the energy distribution of buffer gas atoms (dashed black line).
The deviation of the ion's energy distributions from the Maxwell Boltzmann distribution of the buffer gas atoms results from the additional energy effective potential.
By applying the virial theorem, the pole order dependent energy distribution in the Boltzmann regime is given by
\begin{equation}
P_\mathrm{br}(E_{\mathrm{i}})\propto E_{\mathrm{i}}^{\frac{1}{2}+ \frac{1}{n-1}} \, \exp{(-E_{\mathrm{i}}/k_B T_a)} \, .
\label{eq:BoltzmannE}
\end{equation}
This analytic expression matches well to the results of our numeric simulations.
For large multipole orders, the distribution converges towards the Maxwell Boltzmann distribution of the buffer gas, whereas, for a Paul trap the ion's mean energy exceeds the mean energy of the buffer gas by a factor of $3/2$ (see inset of Fig.~\ref{fig3}).

The ion's spatial distribution is obtained by summing over all radial probability distributions $p_i(r)$ (Eq.~\eqref{eq:piofr}) weighted with the average time the ion spends on this energy shell (Eq.~\eqref{eq:tau}).
The resulting spatial distributions are well described by
\begin{equation}
p_i(r)\propto \exp{\left( - \frac{V_\mathrm{eff}(r)}{k_B T_a} \right)} \, .
\label{eq:BoltzmannR}
\end{equation}
For a Paul trap this corresponds to a Gaussian density distribution, converging towards a boxlike density distribution for $n\rightarrow \infty$.

\begin{figure}[bt]
\includegraphics[width=\columnwidth, trim= 0.3cm 0.2cm 0.3cm 0.1cm]{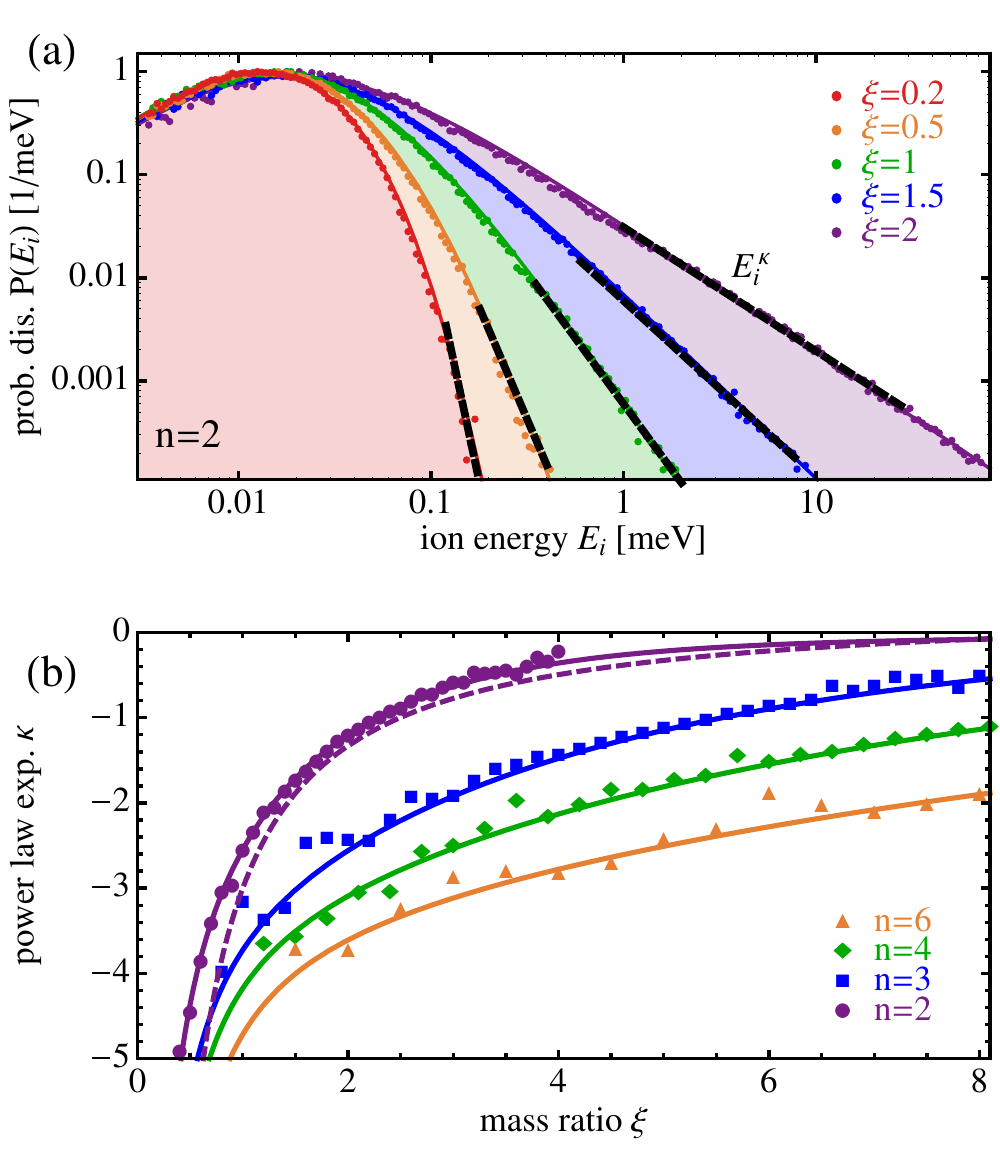}
\caption{
Energy distributions in the power-law regime.
(a) Shown are five distributions corresponding to five different mass ratios.
With increasing mass ratio the energy distribution exhibits a growing power law towards higher energies.
The solid lines correspond to Tsallis functions (Eq.~\eqref{eq:PowerLawE}) which were fitted to the numerical data.
(b) Mass ratio dependence of the power law exponent $\kappa$ for four different multipole orders.
For a fixed mass ratio, the power law exponent is always largest for an ion inside Paul trap and monotonically decreases with increasing multipole order.
The solid lines correspond to Eq.~\ref{eq:kappa1}, the dashed line shows the analytic result of Chen {\em et~al.} \cite{Chen2014} for a Paul trap, which shows reasonable agreement with our numeric results.
}
\label{fig4}
\end{figure}

\subsection{Power-law regime}

With decreasing $\zeta$, the contribution of the micromotion to the atoms effective energy (Eq.~\eqref{eq:effectiveEnergy2}) becomes significant.
It is important to notice that there is a fundamental difference between the two contributions to the atoms effective energy.
On the one hand, the thermal energy has a fixed value and for large ion energies ($E_i \gg \frac{3}{2}k_B T_a$) the probability for the ion to gain energy from the thermal motion of the buffer gas decreases exponentially.
On the other hand, the micromotion induced energy is directly proportional to the ions energy.
Consequently, even at very large energies $E_i \gg \frac{3}{2}k_B T_a$, the probability for the ion to gain energy (as shown in Fig.~\ref{fig0}) from the micromotion remains constant.
As a result, the ion can gain a large fraction (or even multiples) of its energy during every collision which can cause a temporary exponential growth of the ion's energy, if a series of consecutive heating collisions occurs.
It is well known, that such kind of multiplicative processes lead to the emergence of power law distributions \cite{Devoe2009, Zipkes2011}, also known as heavy sided or levy flight distributions \cite{Sakaguchi2001, Tessone2006}.

Fig.~\ref{fig4} (a) shows the resulting energy distributions for different mass ratios in a Paul trap.
The energy distributions in the power-law regime $P_{plr}(E_{\mathrm{i}})$ are well described by Tsallis functions (solid lines) given by \cite{meir2016}

\begin{equation}
P_\mathrm{plr}(E_{\mathrm{i}})\propto \, \frac{E_{\mathrm{i}}^{\frac{1}{2}+ \frac{1}{n-1}}}{\left[ 1+a \, E_{\mathrm{i}}/(k_B T_a) \right]^b} \, ,
\label{eq:PowerLawE}
\end{equation}

which were fitted to the numerical data using the two free parameters $a$ and $b$.
These Tsallis functions are characterized by a power law towards low as well as large energies, centered around the characteristic energy scale $k_B T_a$.
The power law towards low energies is the same as in the Boltzmann regime (Eq.~\eqref{eq:BoltzmannE}).
The power law $E^{\kappa}$ towards higher energies exhibits a steadily rising power law exponent $\kappa$ for increasing mass ratios, as the impact of micromotion induced heating grows.
Fig.~\ref{fig4} (b) shows the mass ratio dependence of $\kappa$ for four different multipole orders.
In lack of an analytical model, the mass ratios is well described by

\begin{equation}
\kappa \approx A \exp \left[ - \xi/B \right] - C \xi^{-D} \qquad (\mathrm{for} \; \kappa \lesssim -0.5)  \; ,
\label{eq:kappa1}
\end{equation}

with the four fit parameters $A= 3.5$, $B= 2.6n-3.8$, $C= n/6+0.5$ and $D= 1.2$ 
If the exponent is close to zero the ion's energy diverges very fast, making it impossible to calculate a steady state energy distribution. 
Already for $\kappa \geq -2$ the ion's mean energy diverges and the ion is no longer confined by the trap.
This condition is commonly used to define the critical mass ratio $\xi_{\mathrm{crit}}$ \cite{Devoe2009,Chen2014}. From the simulations we obtain $\xi_{\mathrm{crit}}\approx 1.4 (n-1)$, which is in good agreement with the analytical result derived before (Eq.~\eqref{eq:xiCrit}).
The critical mass ratio corresponds to a thermalicity parameter of $\zeta \approx 0.6$ for a Paul trap approaching $\zeta \approx 0.5$ with increasing multipole order.

\begin{figure}[tb]
\includegraphics[width=\columnwidth, trim= 0.3cm 0.2cm 0.3cm 0.1cm]{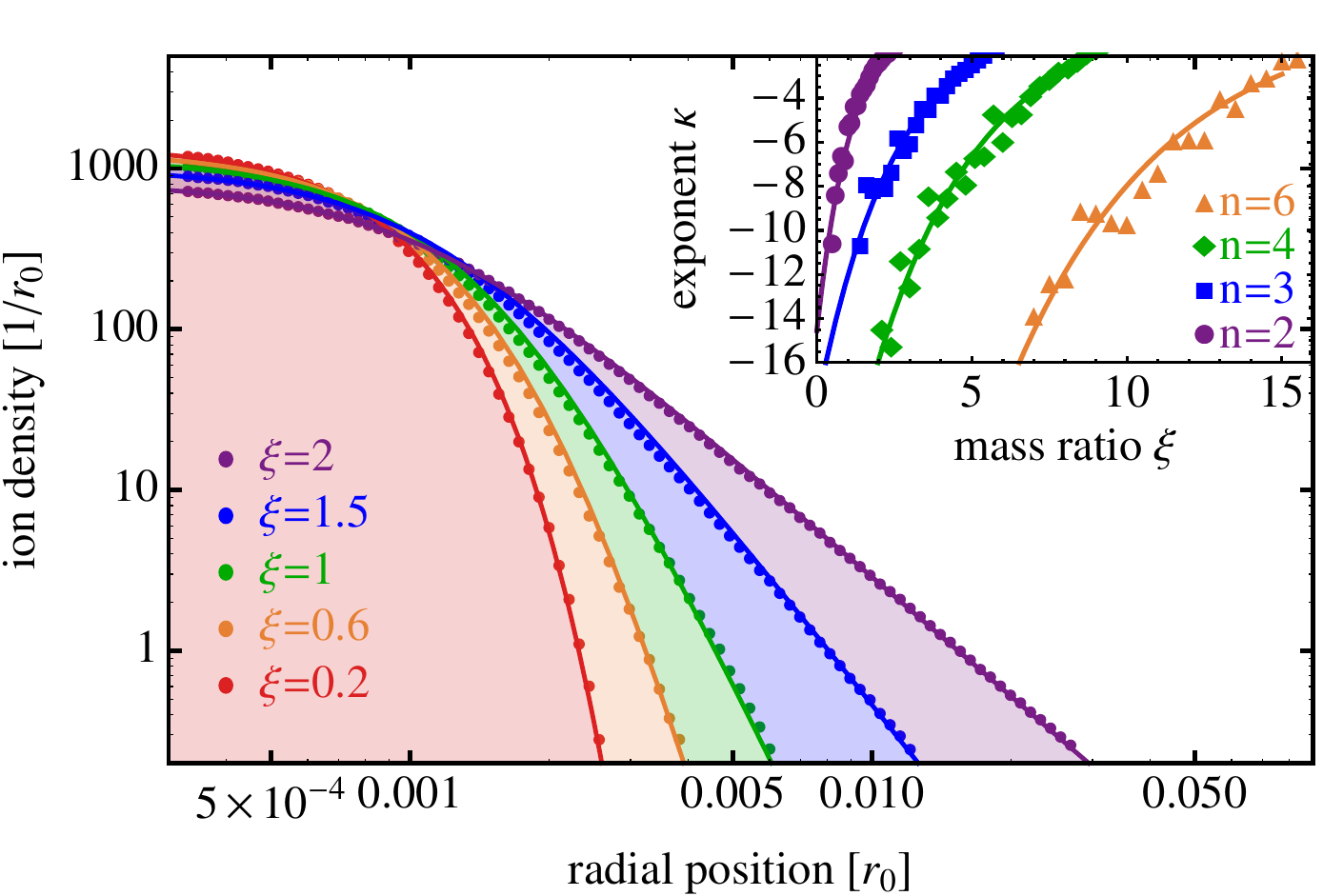}
\caption{
Spatial distribution in the power-law regime. Shown are five distributions for an ion inside a Paul trap, corresponding to different mass ratios. For small mass ratios, the spatial distribution corresponds to the one in the Boltzmann regime (Eq.~\eqref{eq:BoltzmannR}). For larger mass ratios, the distributions exhibit a power law towards large radii, much like the energy distributions. The inset shows the mass ratio dependence of the power-law exponent for different multipole orders. The solid lines are a guide to the eye.
}
\label{fig5}
\end{figure}

Fig.~\ref{fig5} shows the corresponding spatial distributions for a Paul trap, which show a similar power law behavior \cite{Devoe2009}.
The inset shows the exponent of the power law for four different multipole orders.
An ion inside a Paul trap always exhibits the largest power law exponent, which steadily decreases with increasing multipole order.

\begin{figure}[tb]
\includegraphics[width=\columnwidth, trim= 0.3cm 0.2cm 0.3cm 0.1cm]{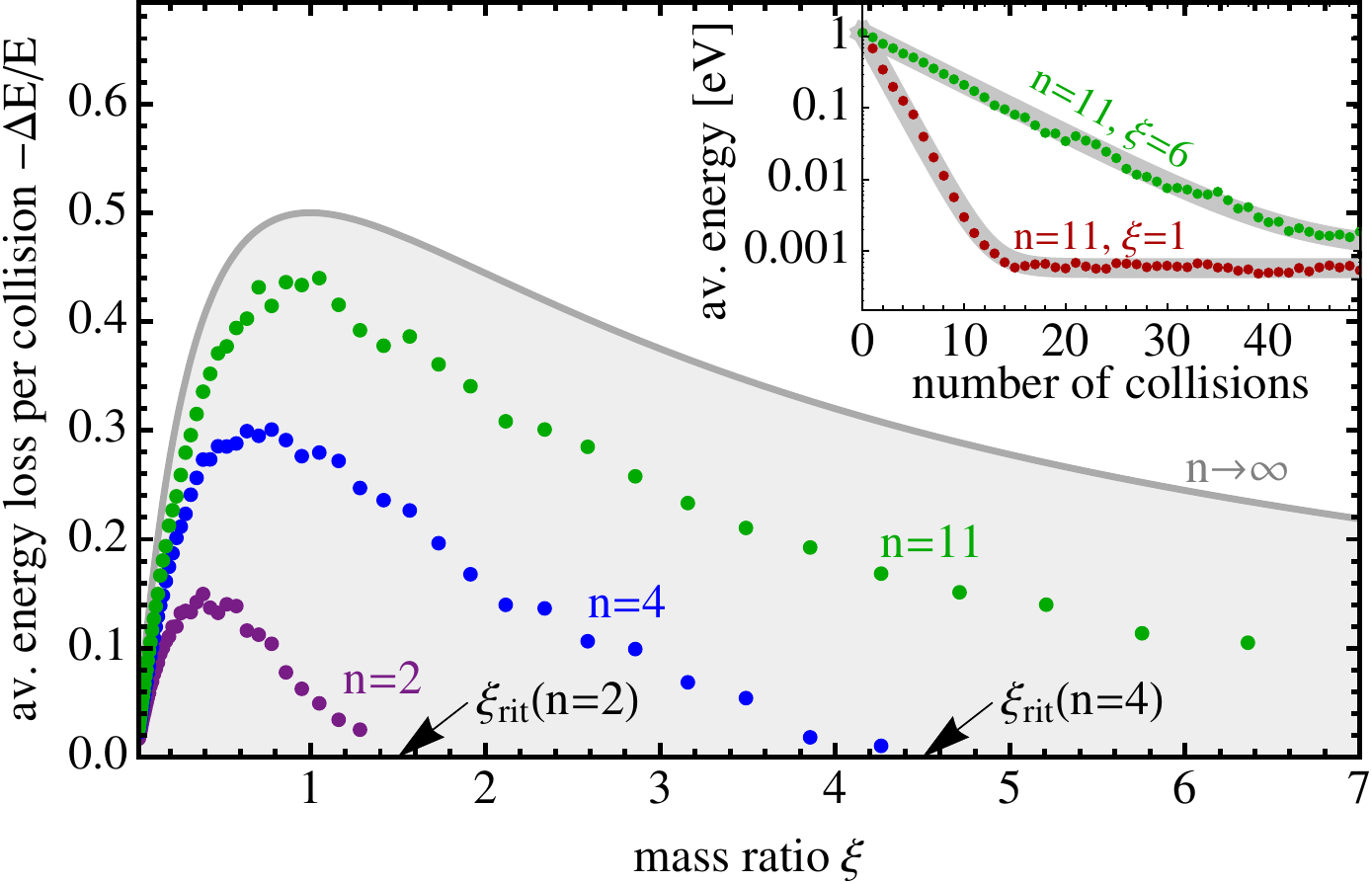}
\caption{
Thermalization rates in different rf traps. Shown is the average energy loss per collision for different mass ratios and multipole orders. The gray curve shows the average energy loss in absence of the rf trap which is equivalent to a trap with an infinite multipole order.
The mass ratio where the average energy loss turns into an energy gain, corresponds to the critical mass ratio.
The inset shows two exemplary energy traces. The average energy loss was extracted from the exponential decay at the beginning of the trace where the ion's energy exceeds the one of the buffer gas.
}
\label{fig6}
\end{figure}

Another important quantity characterizing the sympathetic cooling process is the thermalization rate.
On the one hand, the micromotion induced heating is strongly reduced for small mass ratios.
On the other hand, small mass ratios greatly reduce the average energy transfer per collision.
Fig.~\ref{fig6} shows the resulting average energy loss per collision for different trap orders.
The solid line corresponds to the thermalization of an untrapped ion.
This case represents an upper boundary for the thermalization rate as the presence of rf heating inside the multipole trap results in lower thermalization rates.
In a Paul trap this effect leads to a 3-fold decrease of the maximum cooling rate, which is reached for a mass ratio of $\xi \approx 1/2$.
With growing multipole order, the thermalization rates become larger and the optimal mass ratio tends towards unity as in the case of a free particle.

\subsection{Localization regime}

So far we have focused the discussion on homogeneous spatial distributions of buffer gas filling the entire volume of space.
In this scenario, one finds an upper limit for the atom-to-ion mass ratio, beyond which the ion's are mainly heated until they are expelled from the trap. This limitation can be overcome by localizing the buffer gas to the center of the trap.
We have seen in the previous section, that with decreasing thermalicity parameter, the energy distributions exhibit an increasing power-law tail towards higher energies.
At large energies $E_i$ the probability for a collision to result in a gain of energy (Fig.~\ref{fig0}) increases with the mass ratio.
As the relative energy transfer has a distinct radial dependence, this can be overcome by localizing the buffer gas to the trap center.
As soon as $r_\mathrm{max}$ exceeds the extension of the buffer gas, collisions are restricted to small radii $r/r_\mathrm{max}$.

We find that the resulting energy distributions are bound by an additional exponential decay.
The energy scale of this exponential decay is well estimated by the atoms total effective energy $E_\mathrm{a,tot}$ accessible for collisions with the ion, as given by integrating $\left< E_a \right>$ (Eq.~\ref{eq:effectiveEnergy}) over the entire buffer gas cloud.
For a Gaussian shaped buffer gas cloud with standard deviation $\sigma_a$ this results in an energy of

\begin{equation}
E_\mathrm{a,tot} =  \frac{3}{2} k_\mathrm{B} T_\mathrm{a} + 2^{n-1} (n-1)! \, \xi \, V_\mathrm{eff}(\sigma_\mathrm{a}) \ .
\label{eq:Emax}
\end{equation}

\begin{figure}[bt]
\includegraphics[width=\columnwidth, trim= 0.3cm 0.2cm 0.3cm 0.1cm]{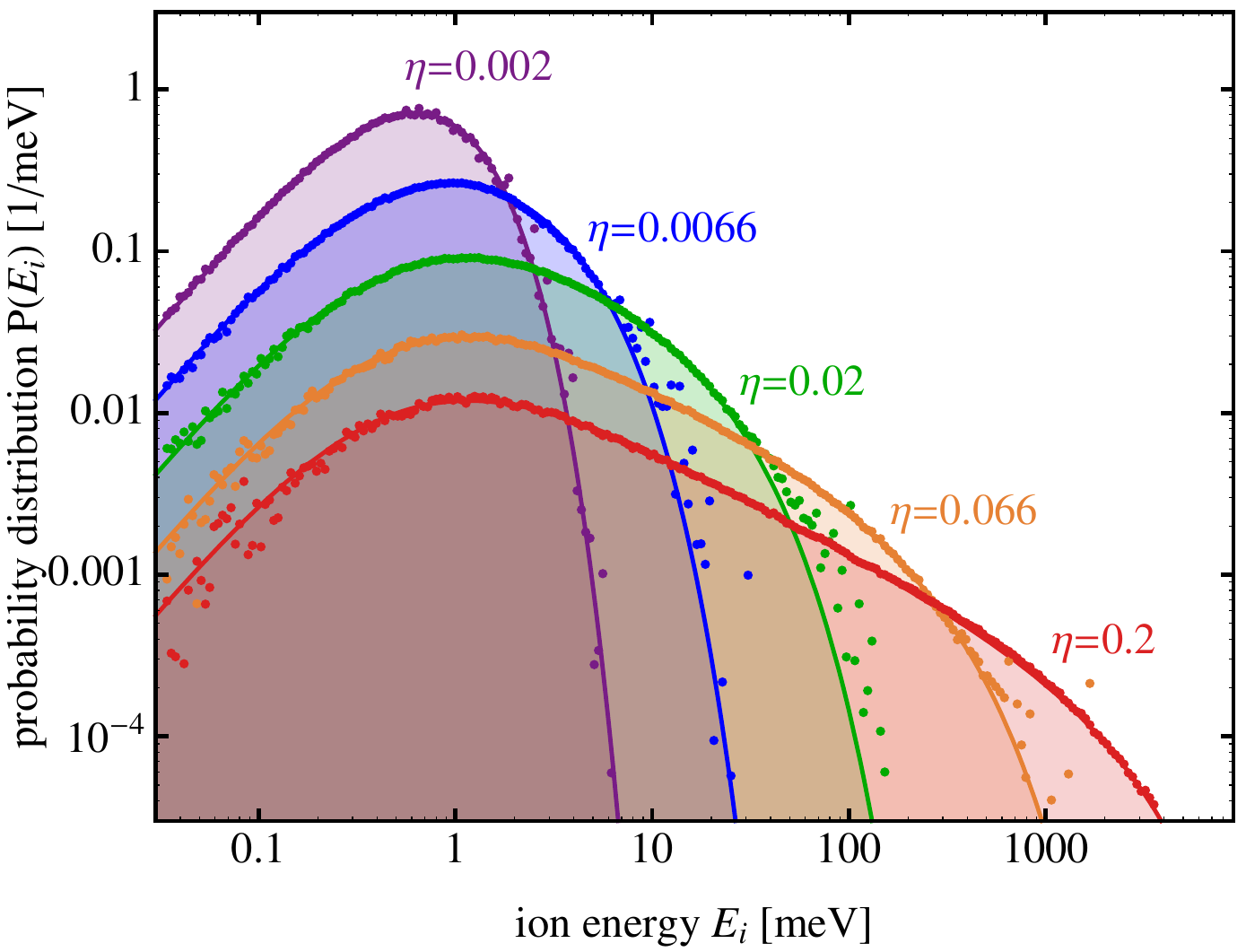}
\caption{
Ion's energy distribution for a localized buffer gas in a Paul trap. The buffer gas has a Gaussian density distribution with standard deviation $\sigma_a = r_0 /10$ and an atom-to-ion mass ratio of $\xi = 3$. The five graphs correspond to five different multipole stability parameters $\eta$ (Eq.~\eqref{eq:stabParameter}).
}
\label{fig7}
\end{figure}

Fig.~\ref{fig7} shows the resulting energy distributions for different stability parameters $\eta$.
With decreasing stability parameter, the energy distribution is bound towards lower energies.
Small stability parameters correspond to a shallow effective potential meaning that the ion probes a volume of the same size as the buffer gas cloud already at relatively low energies $E_i$.
In contrast, for large $\eta$, the ion is strongly confined to the trap center and only at very large energies does the ion experience the localization of the buffer gas.
Instead of varying the stability parameter $\eta$, which can be done by tuning the trap's rf voltage, it is also possible to tune the size of the buffer gas cloud.
Neglecting the contribution of the thermal energy, the atoms total effective energy is proportional to $E_\mathrm{a,tot} \propto \sigma_a^{n-2} U^2$.

\begin{figure}[bt]
\includegraphics[width=\columnwidth, trim= 0.3cm 0.2cm 0.3cm 0.1cm]{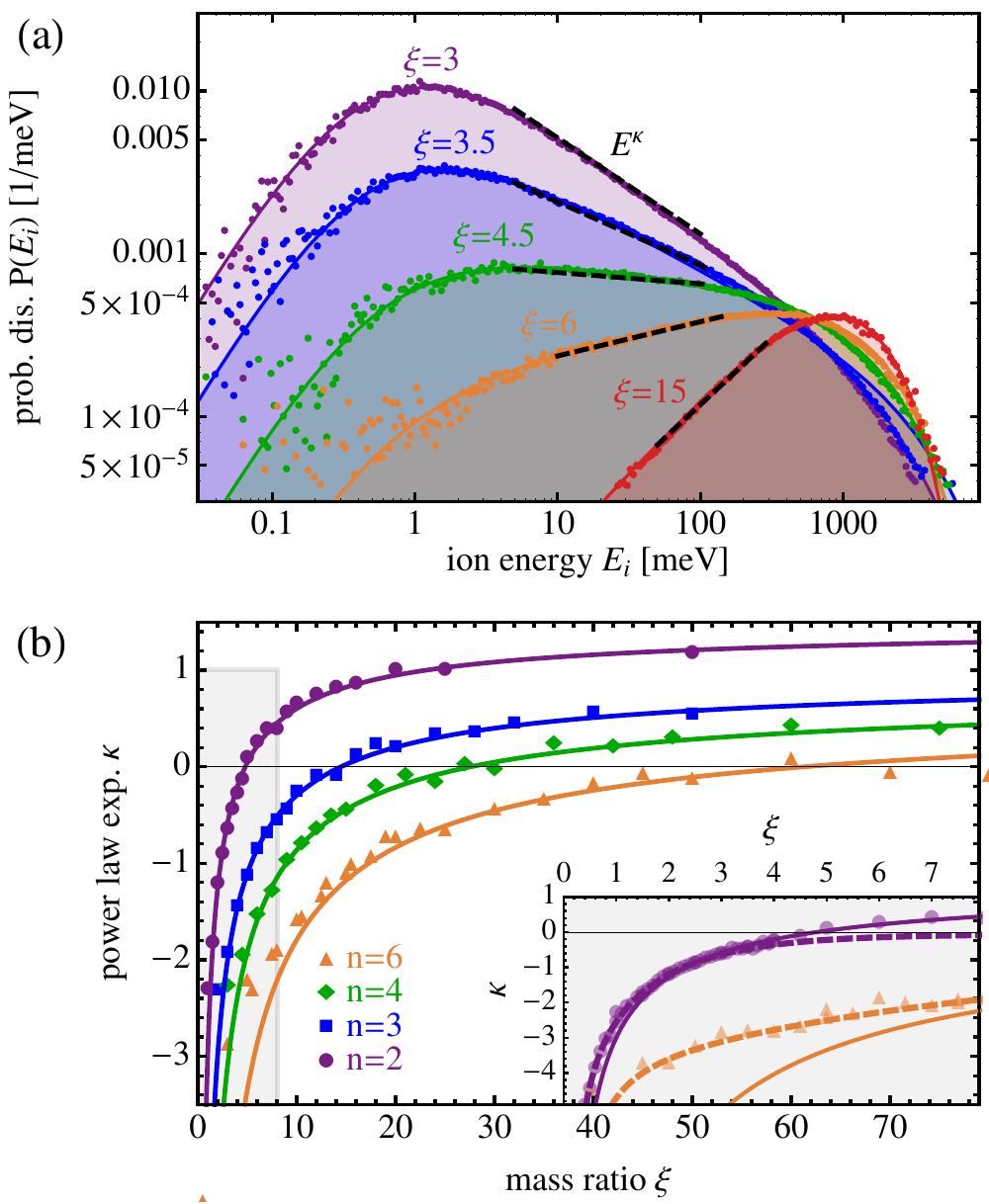}
\caption{
Energy distributions in the localization regime. (a) Ion's energy distribution for different mass ratios in a Paul trap. The energy scale $E_\mathrm{a,tot}$ was kept constant at 1eV. (b) Power law exponents for different multipole orders as indicated by the dashed lines in the upper graph. Shown are the power law exponents as a function of the mass ratio. The solid curves correspond to the analytic expression given in Eq.~\eqref{eq:kappa} which show good agreement with the numerical data for large mass ratios. For small mass ratios however, the exponent is better described by Eq.~\eqref{eq:kappa1} as indicated by the close up into the area marked in gray. Shown is the numerical data for $n=2,6$ and the corresponding analytic expressions from Eq.~\eqref{eq:kappa} (solid lines) and Eq.~\eqref{eq:kappa1} (dashed lines).
}
\label{fig8}
\end{figure}

The energy distribution in the localzation regime is given by
\begin{equation}
P_{lr}(E_{\mathrm{i}})\propto P_{plr}(E_{\mathrm{i}}) \exp{\left[ - \left( \frac{E_i}{\gamma E_\mathrm{a,tot}} \right)^{1/(n-1)} \right]} \, ,
\label{eq:LocalizationE}
\end{equation}
with $\gamma$ being a free parameter which is on the order of one.

The emergence of the additional exponential decay towards large energies also allows the use of heavy buffer gases with $\xi \gg \xi_\mathrm{crit}$.
Fig.~\ref{fig8} shows the energy distributions for different mass ratios with the energy scale $E_\mathrm{a,tot}$ kept constant.
For small mass ratios, the distributions are equivalent to the ones shown in Fig.~\ref{fig7}, with the power law turning into an exponential decay around $E_\mathrm{a,tot}$.
With increasing mass ratio, the power law exponent keeps growing until for $\xi \approx 5$ a plateau-like distribution is reached which has a constant probability all the way from the thermal energy of the buffer gas up to the energy $E_\mathrm{a,tot}$.
For even larger mass ratio, the power law exponent becomes positive, eventually making the thermal energy irrelevant and the energy distribution is fully characterized by $E_\mathrm{a,tot}$.

Fig.~\ref{fig8} shows the power law exponent $\kappa$ as a function of the mass ratio, as obtained from our numerical simulations.
For large mass ratios, the exponent is well described by
\begin{equation}
\kappa \approx A' - C' \xi^{-D'} \qquad (\mathrm{for} \; \kappa\gtrsim -0.5) \, ,
\label{eq:kappa}
\end{equation}
with the fitting parameters $A'=0.5+(n-1)^{-1}$, $C'=2n+0.5$ and $D'=0.7$.
For very large mass ratios the power law converges to the same one found in the Boltzmann regime.
Fig.~\ref{fig8} (b) shows the mass ratio dependence of $\kappa$, with the inset showing a close up for small mass ratios where Eq.~\eqref{eq:kappa} (solid lines) looses its validity and Eq.~\eqref{eq:kappa1} (dashed lines) provides the correct mass ratio dependence.
The transition between the two dependencies happens around $\kappa \approx -0.5$.

\section{Forced sympathetic cooling}

\begin{figure}[tb]
\includegraphics[width=\columnwidth, trim= 0.3cm 0.2cm 0.3cm 0.1cm]{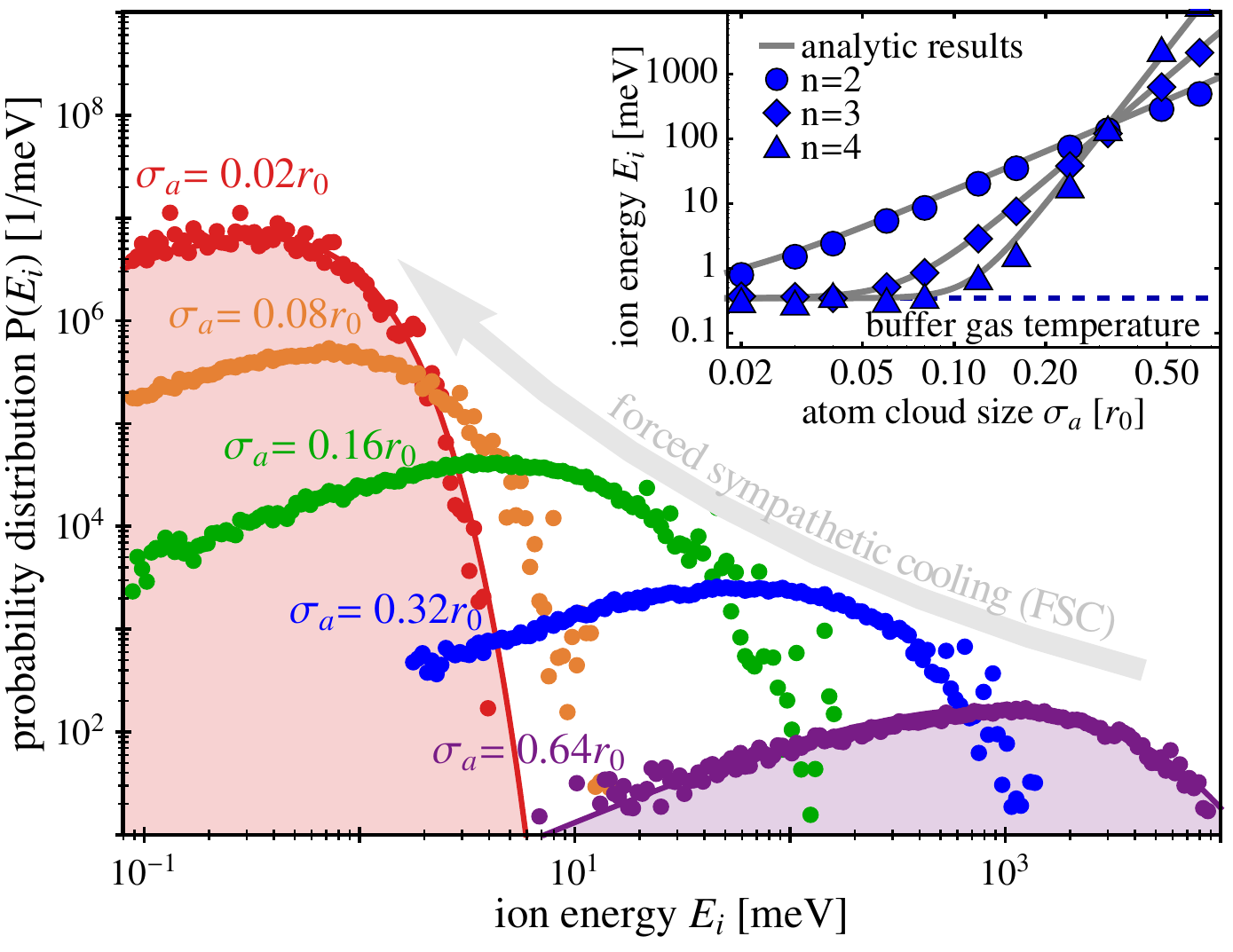}
\caption{
Forced Sympathetic Cooling. Shown are five energy distributions in an octupole trap $(n=4)$ for different buffer gas sizes.
The buffer gas has a temperature of $T_a=4$K and $\xi=30$.
The two curves correspond to the analytic expressions found in the text.
The inset shows the ion's mean energy and the effective buffer gas energy (Eq.~\eqref{eq:Emax}) as a function of the buffer gas size for three different trap orders.
Both energies are in good agreement.
}
\label{fig10}
\end{figure}
The idea behind forced sympathetic cooling (FSC) is to tune the ion's energy distribution by changing the energy scale $E_\mathrm{a,tot} \propto \sigma_a^{2n-2} U^2$. 
FSC does not work for homogeneous buffer gas distributions, as it is the localization of the buffer gas which introduces the energy scale $E_\mathrm{a,tot}$.
In the three regimes discussed above, FSC has different effects on the ion's energy distribution.
In the Boltzmann regime, the impact of FSC is very limited as the characteristic energy scale of the ion's energy distribution is set by the thermal energy of the buffer gas.
Nevertheless, if the energy scale $E_\mathrm{a,tot}$ is chosen smaller than the thermal energy of the buffer gas, small collision radii $r/r_\mathrm{max}$ will be favored.
Eventually, for $E_\mathrm{a,tot} \ll k_B T_a$ the ion's energy distributions independently of the mass ratio, will exactly equal the one of the buffer gas (dashed line in Fig.~\ref{fig3}).
For larger mass ratios, the localization leads to the additional exponential decay towards higher energies as discussed in the context of Fig.~\ref{fig7}.
In this case, FSC can be used to substantially reduce the ion's mean energy as illustrated in Fig.~\ref{fig7} for $\xi=3$.

The largest effect of FSC is achieved for very large mass ratios, where the energy distribution is dominated by the energy scale $E_\mathrm{a,tot}$.
In this case, the entire distribution can be shifted to smaller energies, not just the exponential cut off of the power law.
This is illustrated in Fig.~\ref{fig10} which shows the ion's energy distribution for five different buffer gas sizes in an octupole trap $(n=4)$.
In this case, reducing the buffer gas size by a factor of two leads to a reduction of the ion's mean energy by a factor of 64.
This illustrates that especially in traps with high multipole orders, already small reductions of the buffer gas size can have a large impact on the ion's energy distribution.
Adiabatically changing the size of the buffer gas or the rf voltage does not change the relative overlap of atom and ion distribution and thus the collision rate.
The size of ion and atom cloud always maintains a fixed ratio.

\section{Conclusion}
\label{sec:3}

We have presented a comprehensive model providing an intuitive picture of collisions in an RF trap based on a favorable frame transformation, where the micromotion is assigned to the neutral buffer gas.
In contrast to previous investigations, our model allows the description of an ion inside a trap of arbitrary multipole order colliding with a buffer gas atoms with an arbitrary spatial distribution.
Based on this model we have numerically determined the ion's steady-state energy and spatial distributions. 
Depending on the thermalicity of the buffer gas, as expressed by the parameter $\zeta$, we found three distinct dynamical regimes, characterized by analytical expressions for the ion's equilibrium energy distribution.

For $\zeta \approx 1$ the ion thermalizes with the buffer gas, with small deviations being caused by the confining effective potential.
With decreasing $\zeta$, the energy as well as spatial distribution exhibits a power law towards larger energies/radii.
We found heuristic expressions for the mass ratio dependence of the power law exponent and the critical mass ratio.
In a homogeneous buffer gas the ion cannot stably be trapped for mass ratios much larger than the critical mass ratio which corresponds to a thermalicity parameter of $\zeta \simeq 0.5$.

Using a localized buffer gas however, we found the emergence of a novel regime of stable cooling for ions far beyond the critical mass ratio.
In this regime one can actively tune the ions temperature by controlling the buffer gas' extension and/or the RF-trap fields (forced sympathetic cooling).

Our findings are directly applicable to cooling of ions with laser cooled atoms or He buffer gas in Paul traps (as used in the quantum information and quantum simulation communities) or multipole traps (as used in the chemical reaction and astrochemistry communities). For experiments investigating interactions of ions  with an ensemble of ultracold atoms, the prospect of using heavier atom species opens a whole new range of possible systems which have not been studied yet. We are currently performing experiments on sympathetic cooling of anions with ultracold alkali atoms, using the combination of OH$^-$ in an octupole rf trap interacting with laser cooled and trapped Rb \cite{Deiglmayr2012}. 
 
\section*{Acknowledgments}
This work is supported in part by the Heidelberg Center for Quantum Dynamics and the BMBF under contract number 05P12VHFA6.
B.H. acknowledges support by HGSHire and P.W. by the Studienstiftung des deutschen Volkes.
We thank S. Whitlock, J. Evers, R. Wester and P. Schmelcher for fruitful discussions.


\bibliography{ionatominteractions}

\end{document}